\documentclass[twocolumn,showpacs,showkeys]{revtex4}
\usepackage[utf8]{inputenc}
\usepackage[english]{babel}
\usepackage{amssymb}
\usepackage{amsmath}
\usepackage{graphicx}

\begin{document}

\title{Analytical model for electromagnetic cascades in rotating
electric field}

\author{ E.~N.~Nerush}
\email{nerush@appl.sci-nnov.ru}
\author{ V.~F.~Bashmakov}
\author{I.~Yu.~Kostyukov}
\affiliation{Institute of Applied Physics, Russian Academy of Sciences, 603950 Nizhny Novgorod, Russia}

\begin{abstract}
Electromagnetic cascades attract a lot of attention as an important
QED
effect that will reveal itself in various electromagnetic
field configurations at ultrahigh intensities.
We study cascade dynamics in rotating electric field analytically
and numerically. The kinetic equations for the electron-positron
plasma and gamma-quanta are
formulated. The scaling laws are derived and analyzed. For the
cascades arising far above the threshold the dependence of the
cascade parameters
on the field frequency is derived.
The spectra of high-energy cascade particles are calculated.
The analytical results are verified by numerical simulations.
\end{abstract}

\pacs{12.20.-m, 42.50.Ct, 52.27.Ep, 52.25.Dg}
\keywords{electromagnetic cascades, strong laser field,
kinetic equations, Monte Carlo simulations}
\maketitle

\section{Introduction}

Quantum electrodynamics (QED) effects in a strong
electromagnetic field attract a lot of attention \cite{Mourou2006,Marklund2006}.  
In the last decade a new surge of interest in QED effects in superstrong electromagnetic
field arises thanks to substantial progress in laser technologies
\cite{Yanovsky2008,ELI,HiPER}.  Theoreticians are considering such
effects as vacuum polarization in laser field \cite{King2010},
electron-positron ($e^+e^-$) pair production from vacuum in the field
of colliding laser pulses \cite{BulanovMultiple2010} and in combined
Coulomb and strong laser field \cite{PiazzaCoulomb2010} etc. 

Up to now the electromagnetic fields are not so intense in laboratory to observe
strong-field QED effects directly. However, the interaction of 
ultra-high energy electron beams with electromagnetic fields can be used 
to study the effects as in the frame of the relativistic electron 
the electromagnetic field can be very strong.
The spectra of the scattered electrons and photons arising from the nonlinear Compton scattering 
of intense laser pulse by  $46.6$~GeV electron beam have been measured experimentally
 \cite{Burke1997,Bamber1999}.  The $e^+e^-$ pair production has been also observed.
The strong-field QED effects like channeling radiation 
 and $e^{+}e^{-}$ pair production  at the interaction of 
 relativistic electron beam with crystalline fields
 has been also studied experimentally and theoretically 
 \cite{Baier1998,Uggerhoj2005}. A number of theoretical and numerical models for QED 
 effects accompanying an interaction of relativistic electrons
with strong electromagnetic fields has been developed. 
The evolution of the distribution functions of electrons, positrons and hard photons 
in QED-strong pulsed laser fields interacting with electron beams were
found numerically \cite{Sokolov2010}. In Ref.~\cite{Hu2010} multiphoton trident pair
production in strong laser field was studied in a manifestly
nonperturbative domain in the periodic plane wave approximation. In addition, the energy spectrum of
positrons produced in collision of strong laser pulse with high-energy
electron beam was calculated. The trident pair production amplitude in
a strong laser background was also calculated in
Ref.~\cite{Ilderton2011}. As opposed to \cite{Hu2010}, the finite
pulse durations were allowed, while the laser fields were still treating
nonperturbatively in strong-field QED. The individual contributions of
the one-step and two-step processes were explicitely identified.
 
Another effect that can be very important at high intensities is the
generation of electromagnetic cascades \cite{Bell2008,Fedotov2010}.
The cascades develop as follows. A slow electron can be accelerated
in superstrong laser fields up to very high energy. Then it can
produce a high-energy photon $\gamma$ by Compton scattering with $n$
laser photons involved in the process:
\begin{equation}
e + n \omega \rightarrow e' + \gamma.
\label{Compton}
\end{equation}
The resulting photon creates a pair in a photon-multiphoton
collision:
\begin{equation}
\gamma + n \omega \rightarrow e^+e^-.
\label{BreitWheeler}
\end{equation}
This reaction can be considered as the strong-field generalization of
the Breit--Wheeler process \cite{Marklund2006,Breit1934}. The produced electron
and positron as well as the initial (seed) electron are accelerated by the
field and become capable to produce new high-energy photons by the
process \eqref{Compton}. Hence, the number of the particles grows
rapidly due to the deposition of the laser energy into the particle
acceleration. As a result, the relativistic electron-positron-gamma plasma
is produced.
Besides pair creation by consecutive processes \eqref{Compton} and
\eqref{BreitWheeler}, the particle multiplication can occur by the
process of Bethe--Heitler type (or trident process)
\cite{Marklund2006,Bethe1934}: $e + n\omega \rightarrow e'e^+e^-$. 
However, the
threshold intensity for cascade development is about $I \sim 10^{25}
\text{ W}/\text{cm}^2$, and for modern laser system the characteristic wavelength is
about $\lambda_l \sim 1 \text{ }\mu\text{m}$. At such parameters the
overall probability of processes \eqref{Compton} and
\eqref{BreitWheeler} are much higher than the probability of the
Bethe--Heitler process.

The seed $e^+ e^-$ pair can be created as a result of the instability 
of QED vacuum in strong electromagnetic field \cite{BulanovMultiple2010,Sauter1931,Heisenberg1936,Schwinger1951}.
The probability of the vacuum pair creation increases in the field of two colliding
laser pulses \cite{Brezin1970,Avetissian2002} or in the focused laser 
pulse \cite{Narozhny2004}. 
 However, the threshold intensity for the pair creation by
 colliding and focused laser pulses with $\lambda_l \simeq 1 \text{ }\mu\text{m}$
is very high and can be up to  $2.3 \times 10^{26} \text{ W}/\text{cm}^2 $ 
for the pulse duration $10 \text{ fs}$ and circular polarization \cite{Narozhny2006}.
The use of high-energy photons as a seed for the cascade is more preferable
than that of free electrons and positrons because the ponderomotive force 
of the laser pulses pushes the charged particles out from the high-intensity 
region. The seed particles can be also produced by  atom
ionization \cite{Hu2002} or pair production in the superposition of
laser field and the Coulomb field of heavy nuclei
\cite{PiazzaCoulomb2010}. When the number of 
the created cascade particles becomes large the plasma effects 
are important and the significant amount of the laser energy
can be absorbed due to production and heating of the
electron-positron pair plasma. This can limit the attainable intensity of
high-power lasers \cite{Fedotov2010,PRL2011}.

The dimensionless parameters characterizing QED effects can
be introduced \cite{Nikishev1985,Berestetskii1982}:
\begin{eqnarray}
a_0 = \frac{e \sqrt{-A_\mu A^\mu }}{m c }, \\ 
\chi_{e,\gamma} = \frac{e\hbar}{ m^3 c^4 } \left| F_{\mu \nu} p^{\nu}
\right| \simeq \frac{ \left( {\mathbf E} + [ {\mathbf v}/c, {\mathbf
B} ] \right)_{\perp}}{ E_{cr} } \gamma,
\end{eqnarray}
where $A_\mu$ is the $4-$vector of the field potential,
 $\mathbf E$ and $\mathbf B$ are the electric and magnetic field
strength, respectively, $F_{\mu \nu}$ is the field-strength tensor,
$p^{\nu}$ is the particle four-momentum, $\gamma$ is the particle
Lorentz factor ($\gamma = \hbar \omega/mc^2$ for the photon), $\mathbf
v$ is the particle velocity, $E_{cr} = m^2 c^3/e \hbar$ is the
QED critical field strength, $e>0$ is the value of electron
charge, $m$ is the electron mass, $c$ is the speed of light, $\hbar$
is the Planck constant, $\omega$ is the photon frequency, 
$\omega_l$ is the laser frequency, symbol
$\perp$ denotes the component which is perpendicular to the particle
velocity. The first parameter $a_0$ characterizes the laser filed strength. 
If the laser field is week $a_0 \ll 1$ then mostly $n = 1$ in 
Eqs.~\eqref{Compton} and \eqref{BreitWheeler}. In the opposite limit $a_0 \gg 1$ the 
electron absorbs a large number of photons and the radiation 
spectrum becomes synchrotron-like \cite{Nikishev1985}. 
The second parameter $\chi _e $ characterizes the regime of the photon 
emission. In the limit $\chi _e \ll 1$ the radiation can be described 
classically (except the case when the number of the emitted photons is
small, see Ref.~\cite{NIMA}) while in the opposite limit the recoil
imposed on the electron by the emitted photon is substantial and the
quantum approach should be used. 
Similarly, pair production is significant in the limit $\chi_{\gamma} \gg 1$
while it is exponentially suppressed in the classical limit $\chi_{\gamma} \ll 1$.

At high intensity the so-called radiation formation length
\cite{Berestetskii1982} becomes much smaller than the laser wavelength
and the characteristic path passed by the electron between two
consecutive photon emissions.
This regime is often referred as quasistatic or tunneling
and can be treated in the framework of quasiclassical approach  \cite{Baier1998,Berestetskii1982}. 
The Compton scattering is treated just as the transition of the electron from one level to another in
external constant and homogeneous electromagnetic field that is accompanied by photon emission.
The pair production is considered similarly.  In addition,
the quasiclassical approach implies that the particles moves along the classical
trajectories between the instances of photon emission or
pair production that strongly simplifies calculations. 

Quantum dynamics of the particles and their interaction 
are complex phenomena and a number of numerical models is proposed.
There are numerical models based on the first principles
like solving Dirac's equation and calculating the cross-sections 
of QED processes \cite{Keitel2010}. To simulate collective dynamics 
of the particles the numerical solution of the kinetic equations
can be used \cite{Sokolov2010}. Another probably more numerically efficient 
approach is provided by Monte Carlo methods.
A number of sophisticated models like well-known Geant4 toolkit for 
the simulation of the passage of particles through matter \cite{Geant4}
are based on Monte Carlo methods. Monte Carlo calculations are also
used to analyze the cascade development in the strong electromagnetic 
fields \cite{Anguelov1999,SokolovArxiv}.
Recently particle-in-cell methods have been also used to simulate the 
effect of the self-generated electromagnetic fields of the produced $e^+e^-$ pairs
on the cascade development \cite{Timokhin2010,PRL2011}.
 
In this paper we examine and solve analytically the kinetic equations for the particle
distribution functions in the electromagnetic cascade in the homogeneous rotating
electric field. This is one of the most simple field configuration
where electromagnetic cascades arise. Besides, the rotating electric field can be
found in the $B$-node of the circularly polarized standing wave 
formed by two co-propagating circularly polarized laser pulses.
The plane $B=0$ is the most favorable for the cascade development.
However, for the simplicity we consider homogeneous electric field.
In order to analyze the particle distribution we use the following
assumptions: (i) we use 
the quasiclassical synchrotron formulae for the  probabilities of photon emission and 
pair production, and neglect the angles $\lesssim 1/\gamma$; (ii) the probability rates are averaged over 
the polarization of the particles; (iii) the 
self-generated electromagnetic fields of the particles produced 
in the cascade are neglected. 
The last assumption is valid at the initial stages
of the cascade development when the density 
of the self-generated electron-positron pair plasma is not high \cite{PRL2011}.
Under the assumption that the number of the
cascade particles doubles many times on the period of the field
rotation the analytical solutions for the
distribution functions are obtained for the region in the phase space
$\gamma \gg \langle \gamma \rangle$ and $\chi \gg 1$, where $\langle
\gamma \rangle$ is the average relativistic factor of the cascade
particles.
The obtained solution is used to calculate the energy spectra of the electrons, the positrons and
the photons. The analytical results are compared with the results of the numerical simulations 
performed using the particle-in-cell Monte Carlo technique described in
Ref.~\cite{PRL2011}.

The paper is organized as follows. In Sec.~II we consider the
electron, positron and photon motion in the rotating electric field. 
In Sec.~III the cascade kinetic equations for the particle
distribution functions are discussed. The cascade scalings are derived
and discussed in Sec. IV and Sec. V. The
distribution functions of the cascade particles are
derived in Sec.~VI. In Sec.~VII the obtained
results are verified by the numerical simulations and the asymptotic
formulae for the particle spectra are given in Sec.~VIII. Finally, in
Sec.~IX summary and discussion are given.

\section{Particle motion in the absence of photon emission and pair production}

\begin{figure*}
\includegraphics[]{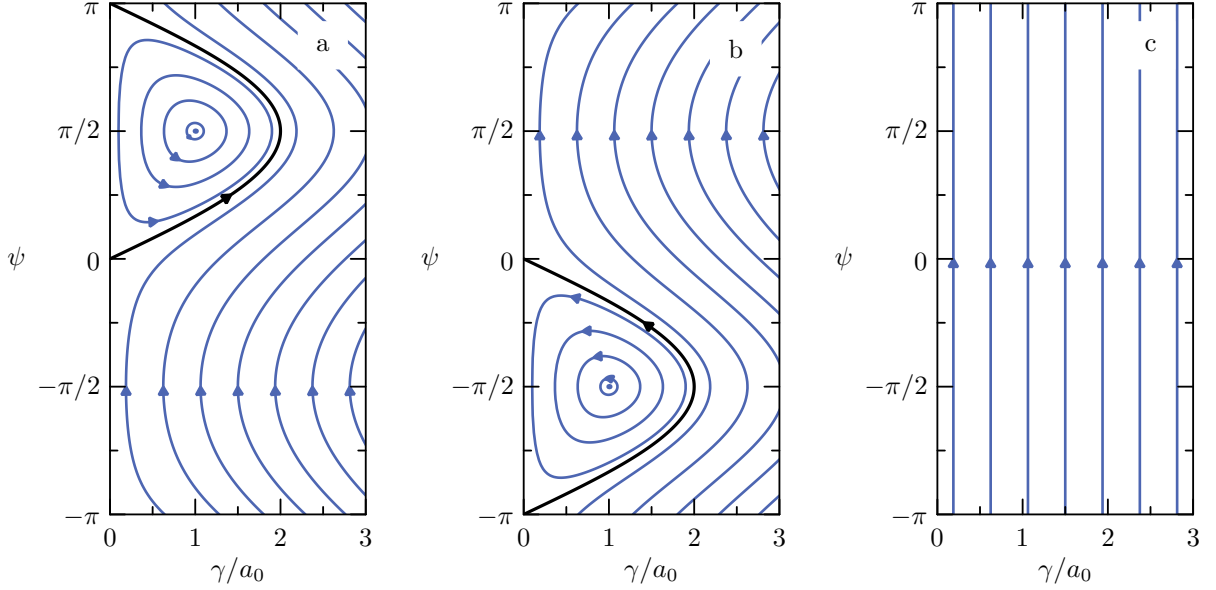}
\caption{The trajectories in $p-\psi$ space for (a) electrons, (b)
positrons and (c) photons moving in the rotating electric field. Photon
emission and pair production are not taken into account. The bold
lines correspond to the trajectories which separate bounded and open
trajectories.}
\label{fig_phasespace}
\end{figure*}

First we study particle motion in the rotating electric field in the framework of the classical approach, 
neglecting photon emission and $e^+e^-$ pair production. 
The field components are the following:
\begin{eqnarray}
E_x = E_0 \cos(t+\phi_0), \label{Ex} \\
E_y = E_0 \sin(t+\phi_0), \label{Ey}
\end{eqnarray}
that leads to the following equations of motion
\begin{eqnarray}
\frac{ dp_x}{ dt} = -\cos(t+\phi_0), \label{dpxdt} \\
\frac{ dp_y}{dt} = -\sin(t+\phi_0), \label{dpydt}
\end{eqnarray}
where $x$, $y$ are the Cartesian coordinates normalized to $c/\omega_l$ in the plane of the
rotation of the vector $\mathbf E$, $t$
normalized to $1/\omega_l$,
$\mathbf p$ is the electron momentum normalized to $mc a_0 =
eE_0/\omega_l$, and
$\phi_0$ is initial phase. 
We are interested only in particle dynamics in the momentum space 
so far as the electric 
field is homogeneous. We introduce $\psi$ universally for all
species of the particles as the angle between 
$-{\mathbf E}$ and $\mathbf p$. Small and positive
values of $\psi$ mean that particle lags behind the vector $-\mathbf
E$ in the rotation. Then we have: $p_x = -p \cos( t+\phi_0 -\psi)$, $p_y = -p
\sin(t+\phi_0 - \psi)$, and  Eqs.~\eqref{dpxdt} and \eqref{dpydt} can be rewritten 
as follows
\begin{eqnarray}
\frac{d\psi}{ dt} = 1 - \frac{\sin \psi}{ p}, \label{dpsidt} \\
\frac{dp}{ dt} = \cos \psi.
\label{dpdt}
\end{eqnarray}
These equations can be reduced to the single equation in the plane $p-\psi$:
\begin{equation}
\frac{d (\sin \psi)}{ dp } = 1 - \frac{ \sin \psi }{ p},
\end{equation}
The solution of the equation is the following:
\begin{equation}
\sin \psi = \frac{ p}{2} + \frac{p_0}{p} \left( \sin \psi_0 -
\frac{p_0}{2} \right), \label{phasespace}
\end{equation}
where $p_0$ and $\psi_0$ are the initial values of the
electron momentum and the angle, respectively. The electron trajectories
(\ref{phasespace}) are shown in Fig.~\ref{fig_phasespace}~(a).
The positron trajectories (Fig.~\ref{fig_phasespace}~(b)) can be obtained
from Eq.~\eqref{phasespace} by the substitution $\psi \rightarrow \psi
- \pi$. What this means is the positrons are accelerated in the
opposite direction to the electron acceleration direction. The photon
momentum does not change during the propagation, and the angle
$\psi$ for a photon grows linearly in the course of time $\psi =
\psi_0 + t$.
Corresponding trajectories are shown in Fig.~\ref{fig_phasespace}~(c).

The restoration of the particle energy in the cascade can be illustrated
using Fig.~\ref{fig_phasespace}. Let us consider initially immobile
electron. At first, from the point $\psi = 0$, $\gamma = 0$ it is
being accelerated along the bold line in Fig.~\ref{fig_phasespace}~(a)
(separatrix): $p
\approx t$ and $\sin \psi = p/2$. Then
it emits a photon and shifts left: its angle $\psi$ remains the same
and its gamma-factor decreases due to the photon emission. Next
according to the phase flow the electron comes again to the
separatrix. This also explains why the electrons are concentrated near
the separatrix.

As we will see
below, the electron distribution function drops along this
line due to the following reasons. First, some of the electrons emit photons and leave the
separatrix. Second, the acceleration of the electrons that have higher
energy starts earlier in the average than the acceleration of the
electrons with lower energies. The number of particles is increasing
exponentially with time in the cascade, hence, the number of particles that gain
higher energy is lower.

\section{Basic equations}

The cascade kinetic equations have been derived a long time ago to study development
of showers initiated by cosmic rays in the Earth's atmosphere 
\cite{Landau1938,Rossi1953}. 
The shower (avalanche) develops due to the bremsstrahlung and pair production
at the particle interaction with nuclei.
The solution of the equations has been obtained in Ref.~\cite{Landau1938}.
Later the solution method has been extended to include multiphoton 
processes \cite{Khokonov2004} and to study the electromagnetic 
showers in strong magnetic field \cite{Akhiezer1994}. 
The kinetic approach have been used in numerical simulations to
analyze the interaction of the intense laser pulse with the electron
beam \cite{Sokolov2010}. The kinetic equations have been also
formulated for the problem of the cascade development in the laser
field and have been used to derive radiation reaction force
\cite{Poisson1999} in
classical limit in Ref.~\cite{FedotovArxiv}.
However, up to now the kinetic cascade equations have been analytically solved only in one-dimensional cases.
On the contrary, we consider theoretically and numerically the two-dimensional cascade.
Moreover, we will take into account that the overall energy of the
cascade particles is growing and can be much more than the seed
particle energy.

The cascade kinetic equations in arbitrary field can be
written as follows:
\begin{widetext}
\begin{multline}
\partial_t f_{e}({\mathbf r},{\mathbf p}) + \frac{\mathbf p}{ p}
\nabla f_{e}({\mathbf r},{\mathbf p}) + \nabla_{\mathbf p} \{
f_{e}({\mathbf r},{\mathbf p}) \times {\mathbf F_{e}}({\mathbf
r},{\mathbf p}) \} = \int_{{\mathbf p}' \parallel {\mathbf p}}
\frac{{p'}^{D-1}}{p^{D-1}} f_{\gamma} ({\mathbf r},{\mathbf p'}) \tilde w({\mathbf p}'
\rightarrow {\mathbf p}) \, dp' \\ + \int_{{\mathbf p}' \parallel
{\mathbf p}} \frac{{p'}^{D-1}}{p^{D-1}} f_{e}({\mathbf r},{\mathbf p'}) w({\mathbf p}'
\rightarrow {\mathbf p}) \, dp' - \int_{{\mathbf p}' \parallel
{\mathbf p}} f_{e}({\mathbf r},{\mathbf p}) w({\mathbf p}
\rightarrow {\mathbf p}') \, dp',
\label{basic_fe}
\end{multline}
\begin{multline}
\partial_t f_{p}({\mathbf r},{\mathbf p}) + \frac{\mathbf p}{ p}
\nabla f_{p}({\mathbf r},{\mathbf p}) + \nabla_{\mathbf p} \{
f_{p}({\mathbf r},{\mathbf p}) \times {\mathbf F_{p}}({\mathbf
r},{\mathbf p}) \} = \int_{{\mathbf p}' \parallel {\mathbf p}}
\frac{{p'}^{D-1}}{p^{D-1}} f_{\gamma} ({\mathbf r},{\mathbf p'}) \tilde w({\mathbf p}'
\rightarrow {\mathbf p}'-\mathbf{p}) \, dp' \\ + \int_{{\mathbf p}' \parallel
{\mathbf p}} \frac{{p'}^{D-1}}{p^{D-1}} f_{p}({\mathbf r},{\mathbf p'}) w({\mathbf p}'
\rightarrow {\mathbf p}) \, dp' - \int_{{\mathbf p}' \parallel
{\mathbf p}} f_{p}({\mathbf r},{\mathbf p}) w({\mathbf p}
\rightarrow {\mathbf p}') \, dp',
\label{basic_fp}
\end{multline}
\begin{multline}
\partial_t f_{\gamma}({\mathbf r},{\mathbf p}) + \frac{\mathbf p}{p}
\nabla f_{\gamma}({\mathbf r},{\mathbf p}) = -\int_{{\mathbf p}'
\parallel {\mathbf p}} f_{\gamma} ({\mathbf r},{\mathbf p}) \tilde
w({\mathbf p} \rightarrow {\mathbf p}') \, dp' \\ + \int_{{\mathbf p}'
\parallel {\mathbf p}} \frac{{p'}^{D-1}}{p^{D-1}} f_e({\mathbf r},{\mathbf p'}) w({\mathbf p}'
\rightarrow {\mathbf p}'-{\mathbf p}) \, dp' + \int_{{\mathbf p}'
\parallel {\mathbf p}} \frac{{p'}^{D-1}}{p^{D-1}} f_p({\mathbf r},{\mathbf p'}) w({\mathbf p}'
\rightarrow {\mathbf p}'-{\mathbf p}) \, dp',
\label{basic_fph}
\end{multline}
\end{widetext}
where $\mathbf r$ is the particle coordinate normalized to
$c/\omega_l$, $\mathbf F$ is the Lorentz force normalized to
$a_0 mc\omega_l$, $D$ is the space dimension, $f_e$, $f_p$ and $f_{\gamma}$ are the distribution
functions of the electrons, the positrons and the photons, respectively, which are
normalized such that
\begin{equation}
\int f_{e,p,\gamma} ({\mathbf r},{\mathbf p}) \, d^ {D} {\mathbf r} \,
d^{D}
{\mathbf p} = N_{e,p,\gamma},
\end{equation}
where $N_e$, $N_p$, $N_{\gamma}$ is the number of the electrons, the positrons and
the photons, respectively. 
$w({\mathbf p}' \rightarrow {\mathbf p}) dp$ is
the probability in time unit for the electron with momentum ${\mathbf
p}'$ to emit a photon and to switch to the state with the value of
momentum in the interval $(p,p+dp)$ and the direction of the momentum
parallel to ${\mathbf p}'$, $\tilde w({\mathbf p}' \rightarrow
{\mathbf p}) dp$ is the probability in time unit for the photon with
momentum ${\mathbf p}'$ to decay with creation of the electron with
momentum in the interval $(p,p+dp)$ directed parallel to ${\mathbf
p}'$ and the positron with momentum ${\mathbf p}'-{\mathbf p}$. Here
we assume that the photon emission and the pair production occur in
synchrotron regime, and we neglect the angles less or about than
$1/\gamma$. Dependences of $f_{e,p,\gamma}$,
$\mathbf F$ on $t$ and $w$, $\tilde w$ on $\mathbf F$ are not written
for the simplicity.

The RHS of Eqs.~\eqref{basic_fe}-\eqref{basic_fph} describes photon
emission and pair production. Namely, the first term in the RHS of
Eq.~\eqref{basic_fe} characterizes creation of the electrons with momentum
$\mathbf p$ by decay of the photons with momenta $p'>p$, the second and
the third terms respectively describe the increase and decrease of the number
of the electrons with momentum $\mathbf p$ due to photon emission. The
first term in the RHS of Eq.~\eqref{basic_fph} characterizes photon decay,
and the last two terms describe emission of new photons by the electrons
and the positrons. 

The kinetic equations in the rotating electric field~\eqref{Ex},
\eqref{Ey} in the plane geometry ($D=2$) can be written as follows \cite{FedotovArxiv}:
\begin{multline}
\partial_t f_{e,p}({\mathbf p}) \mp \cos(t+\phi_0)\frac{ \partial
f_{e,p}({\mathbf p}) }{\partial p_x} \\ \mp \sin(t+\phi_0) \frac{
\partial f_{e,p}({\mathbf p}) }{ \partial p_y} = \int_{{\mathbf p}'
\parallel {\mathbf p}} \frac{p'}{p} f_{\gamma} ({\mathbf p'}) \tilde w({\mathbf p}'
\rightarrow {\mathbf p}) \, dp' + \\ \int_{{\mathbf p}' \parallel
{\mathbf p}} \frac{p'}{p} f_{e,p}({\mathbf p'}) w({\mathbf p}' \rightarrow {\mathbf
p}) \, dp' - W({\mathbf p}) f_{e,p}({\mathbf p}),
\label{fe}
\end{multline}
\begin{multline}
\partial_t f_{\gamma}({\mathbf p})  = -\tilde W({\mathbf p})
f_{\gamma} ({\mathbf p}) \\
+ \int_{{\mathbf p}' \parallel {\mathbf p}}
\frac{p'}{p} f_e({\mathbf p'}) w({\mathbf p}' \rightarrow {\mathbf p}'-{\mathbf p})
\, dp' \\ + \int_{{\mathbf p}' \parallel {\mathbf p}} \frac{p'}{p} f_p({\mathbf p'})
w({\mathbf p}' \rightarrow {\mathbf p}'-{\mathbf p}) \, dp',
\label{fph}
\end{multline}
where the relation $\tilde w({\mathbf p}' \rightarrow {\mathbf
p}'-{\mathbf p} ) = \tilde w( {\mathbf p}' \rightarrow {\mathbf p} ) $
is taken into account, "$-$" in "$\mp$" corresponds to the equation
for the electron distribution function and "$+$"
corresponds to the equation for the positron distribution function,
\begin{eqnarray}
f_{e,p,\gamma}({\mathbf p}) = \int f_{e,p,\gamma}({\mathbf r},{\mathbf
p}) \, d^2 {\mathbf r}, \\
W({\mathbf p} ) = \int w({\mathbf p} \rightarrow {\mathbf p}') \, dp',
\\
\tilde W( {\mathbf p} ) = \int \tilde w( {\mathbf p} \rightarrow
{\mathbf p}') \, dp'.
\end{eqnarray}

In order to eliminate the dependence of the force and the
probabilities on time
in Eqs.~\eqref{fe} and \eqref{fph}, we make the change of variables
$p_x,\text{}p_y \rightarrow p, \text{ } \psi$ as follows. Introducing
particle distribution functions in new variables
\begin{multline}
g_{e,p,\gamma}(t,p,\psi) = p f_{e,p,\gamma}(t,p_x,p_y) = \\ p
f_{e,p,\gamma}( t, -p\cos( t+\phi_0-\psi),-p\sin( t+\phi_0 - \psi)),
\end{multline}
we obtain:
\begin{equation}
\frac{ \partial g_{e,p,\gamma}}{ \partial t} = p \frac{ \partial
f_{e,p,\gamma}}{ \partial t} - \frac{ \partial g_{e,p,\gamma}}{
\partial \psi},
\end{equation}
\begin{multline}
\cos(t+\phi_0)\frac{ \partial f_{e,p}}{ \partial p_x} + \sin( t+
\phi_0 ) \frac{ \partial f_{e,p}}{ \partial p_y} \\ = \frac{\sin \psi }{
p^2} \frac{ \partial g_{e,p}}{ \partial \psi } - \cos \psi
\frac{ \partial } {\partial p} \frac{ g_{e,p} }{ p}.
\end{multline}
Then, from Eqs.~\eqref{fe} and \eqref{fph} we have:
\begin{multline}
\frac{ \partial g_{e,p} }{ \partial t} = -\frac{ \partial g_{e,p} }{
\partial \psi} \pm \frac{ \sin \psi}{ p} \frac{ \partial g_{e,p}}{
\partial \psi} \mp p\cos \psi \frac{ \partial }{ \partial p} \frac{
g_{e,p}(p, \psi)}{ p} + \\ \int_p^{\infty} g_{\gamma}
(p',\psi) \tilde w(p' \rightarrow p, \psi) dp' + \\ \int_p^{\infty}
g_{e,p}(p',\psi) w(p' \rightarrow p,\psi) \, dp' - W
g_{e,p}, \label{ge_basic}
\end{multline}
\begin{multline}
\frac{\partial g_{\gamma} }{ \partial t} = - \frac{ \partial
g_{\gamma}}{ \partial \psi } - \tilde W
g_{\gamma} + \\ \int_p^{\infty} \left[ g_e(p',\psi)+ g_p(p',\psi)
\right] w(p' \rightarrow p'-p, \psi) \, dp',
\label{gph_basic}
\end{multline}
where $g_{e,p,\gamma}$, $W$ and $\tilde W$ are defined at the point
$(p,\psi)$.

The probability rates $w$, $\tilde w$ contained in
Eqs.~\eqref{ge_basic}, \eqref{gph_basic} can be found in the framework
of the quasiclassical theory \cite{Berestetskii1982, Baier1998}:
\begin{multline}
w( {\mathbf p}' \rightarrow {\mathbf p} ) = - \frac{ \alpha}{ 
\varepsilon_l
a_0 } \frac{1}{ {p'}^2 } \left[ \int_{\varkappa}^{\infty}
\operatorname{Ai} (\xi) \, d\xi  + \right. \\ \left. \left( \frac{2
\eta^{2/3}}{ (1-\eta)^{2/3}} + \frac{ (1-\eta)^{4/3}}{ \eta^{1/3}}
\right) {\chi'}^{2/3} \operatorname{Ai}'(\varkappa) \right],
\label{w_basic}
\end{multline}
\begin{multline}
\tilde w( {\mathbf p}' \rightarrow {\mathbf p} ) = \frac{ \alpha}{
\varepsilon_l a_0 } \frac{1}{{p'}^2} \left[ \int_{\tilde \varkappa}^{\infty}
\operatorname{Ai}(\xi) \, d\xi + \right. {\chi'}^{2/3} \operatorname{Ai}'(\tilde \varkappa) \\ \left. \times \left( 2
\eta^{2/3}(1-\eta)^{2/3} - \frac{1}{ \eta^{1/3} (1-\eta)^{1/3} }
\right)  \right],
\label{tildew_basic}
\end{multline}
where
\begin{eqnarray}
\varkappa = \left( \frac{ 1-\eta }{\eta \chi'} \right)^{2/3}, \label{
kappa} \\
\tilde \varkappa = \left( \frac{1}{ \eta(1- \eta) \chi' }
\right)^{2/3},
\label{tildekappa}
\end{eqnarray}
$\eta = p/p'$, $\chi' = \varepsilon_l a_0^2 p' \sin \psi$ is the quantum
parameter for the initial particle, $\varepsilon_l =\hbar \omega_l/ mc^2$ is
the frequency of the field rotation in Compton units, $\alpha=e^2/\hbar c$ is the fine
structure constant, $\operatorname{Ai}(\xi)$ and $\operatorname{Ai}'(\xi)$ are
the Airy function and its derivative, respectively \cite{Abramowitz}. Eqs.~\eqref{ge_basic}-\eqref{tildew_basic} are the
complete set of equations which describes evolution of the  distribution functions
 in the momentum space for the electromagnetic
cascade in the rotating electric field.

\section{Scalability of the cascade equations}

\begin{figure}
\includegraphics[]{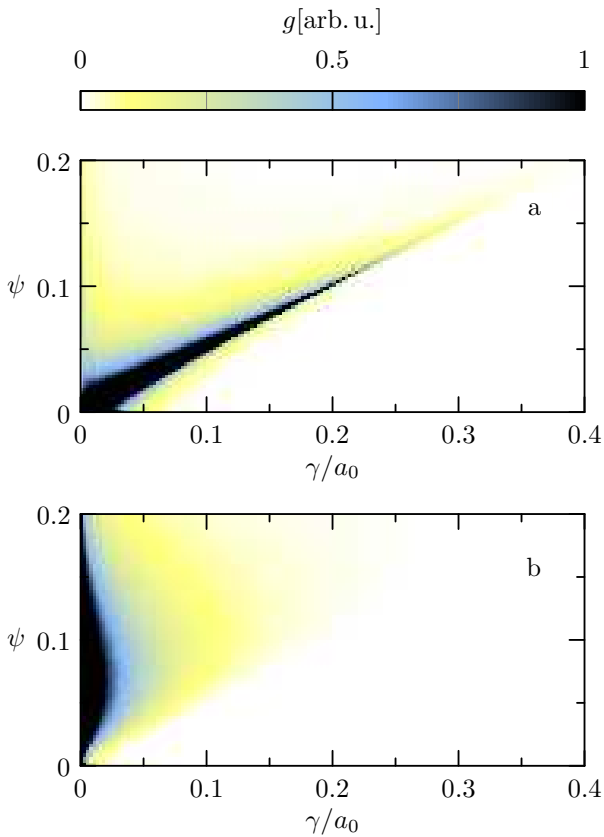}
\caption{The distribution functions of (a) the electrons and (b) the photons
of the electromagnetic cascade developing in the rotating electric field
with the normalized amplitude $a_0 = 1.6\times 10^5$ and $\lambda_l = 0.8
\text{ }\mu\text{m}$.}
\label{fig_spectrum-g}
\end{figure}

Eqs.~\eqref{ge_basic} and \eqref{gph_basic} governing cascade  dynamics depend on two
parameters: period of the electric field rotation and the field strength. However,
it can be shown that if $E_0 \gg \alpha E_{cr}$ the number of the parameters can be reduced.
It is shown in Ref.~\cite{FedotovArxiv} that if $E_0 \gg \alpha
E_{cr}$ and $\lambda_l \gtrsim 2\pi \hbar/mc \alpha^2 \approx 50 \text{ nm}$ the characteristic time between consecutive photon
emissions by the cascade electron (positron) are much smaller than the period of the
electric field rotation $2 \pi/\omega_l$. In other words, the number of the cascade particles
increases in many times during the period of the electric field
rotation, 
 $\Gamma \gg 1$, where $\Gamma $ is the cascade growth rate normalized to $\omega_l$.
 The number of cascade particles increases exponentially  $N \propto \exp ({\Gamma t })$. 

The simplification of the kinetic equations and the reduction of the
number of the governing parameters are possible due to the fact that
in this case the cascade particles are concentrated in the regions
$|\sin \psi| \ll 1$.
As we will see later it follows from numerical simulations that
the condition $|\sin \psi| \ll 1$ holds for the most of the particles
produced in the cascade.
This can be qualitatively explained as follows. Eq.~\eqref{dpdt} implies that $p \sim 1$ 
achieves when the electrons or positrons are accelerated within the 
time interval about the period of the electric field rotation.
If $E_0 \gg \alpha E_{cr}$, the electron emits a lot of photons
within this time interval and loses a great part of the gained energy.
Hence, the electron does not gain high energy from the field and its
characteristic momentum is small $p \ll\ 1$. In accordance with
Eq.~\eqref{dpsidt}, the angle $\psi$ between the electron velocity and the 
direction opposite to the electric field direction, changes very quickly 
until $| \sin \psi | \lesssim p \ll 1$. The photons emitted by the
electrons mostly have $|\sin \psi| \ll 1$. Furthermore, for the
photons $d\psi/ dt = 1$, so they decay before the angle $\psi$ changes
significantly.
Therefore, it is natural to assume that the distribution functions of
the electrons, the positrons and the photons are substantially nonzero only in the
region $| \sin \psi | \ll 1$. 

Using the fact that $|\sin \psi | \ll 1$, the kinetic equations can be
simplified as follows. $|\sin \psi| \ll 1$ if $\psi \approx 0$ or
$\psi \approx \pi$. Let us, for the specificity, consider the region
$| \psi | \ll 1$. In this region the following substitutions can be
made in Eq.~\eqref{ge_basic}: $\cos \psi \rightarrow 1$, $\sin \psi
\rightarrow \psi$.  In the resulting equation as well as in
Eq.~\eqref{gph_basic} (with probabilities from Eqs.~\eqref{w_basic},
\eqref{tildew_basic}) the following substitution
\begin{eqnarray}
t = \beta \hat t, \label{beta_t} \\
p = \beta \hat p, \label{beta_p} \\
\psi = \beta \hat \psi \label{ beta_psi}
\end{eqnarray}
 does not change anything except parameter 
 $\chi \approx \varepsilon_l a_0^2 p \psi = \varepsilon_l a_0^2 \beta^2 \hat
p \hat \psi$, where $\beta $ is arbitrary quantity. Choosing
\begin{equation}
\beta = \frac{1}{ \varepsilon_l^{1/2} a_0 }, \label{beta}
\end{equation}
yields $\chi = \hat p \hat \psi$. The resulting equations for the distribution
functions depend only on the parameter
\begin{equation}
\mu = \frac{ \varepsilon_l a_0 }{ \alpha } = \frac{1}{\alpha} \frac{ E_0 }{
E_{cr} }.
\end{equation}
They do not depend on the frequency $\omega_l$ and, hence, on the
corresponding wavelength
$\lambda_l=2 \pi c/\omega_l$. Thus, the dependence of various cascade characteristics
on $\lambda_l$ is determined only by the scaling
Eqs.~\eqref{beta_t}-\eqref{beta}. For example, it follows from
Eq.~\eqref{beta_t} that for the certain field amplitude and, hence, the
intensity $I=cE_0^2/4\pi$, the growth rates at two different wavelengths
relates as follows: 
\begin{equation} \frac{ \Gamma(I, \lambda_{l,1}
)}{ \Gamma(I, \lambda_{l,2} ) } = \sqrt{ \frac{ \lambda_{l,1}}{
\lambda_{l,2} } }. \label{Gamma_lambda} 
\end{equation} 
This formula is
in good agreement with the results of the numerical simulations as shown in
Sec.~VII.

\begin{figure}
\includegraphics[]{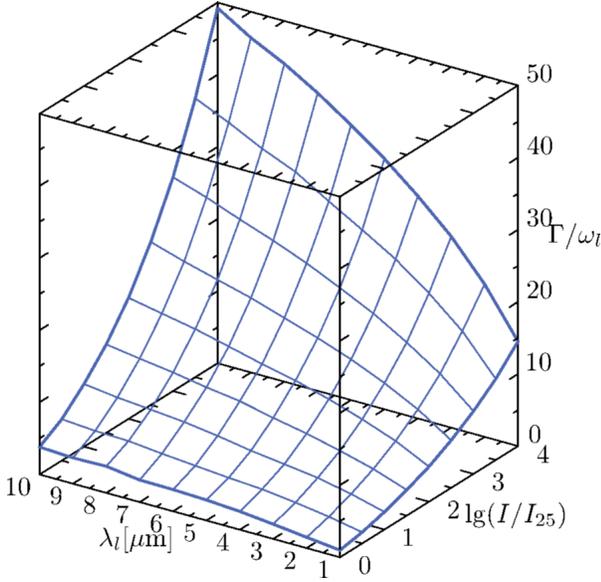}
\caption{The growth rate of the electromagnetic cascade in the rotating
electric field as a function of $I= cE_0^2/4 \pi$ and $\lambda_l = 2\pi
c/ \omega_l$. Here $E_0$ is the amplitude of the electric field,
$\omega_l$ if the cyclic frequency of the field rotation, $I_{25} = 10^{25}
\text{ W}/\text{cm}^2$.}
\label{fig_gamma-surface-plot}
\end{figure}

\section{Scalability of the kinetic equations for particles with $\chi
\gg 1$}

Further simplification can be made if we consider  high-energy "tails" of the
distribution functions with $\chi \gg 1$. This let us to simplify the formulae for the spectral distributions of the
probabilities Eq.~\eqref{w_basic}, \eqref{tildew_basic}. Fist, in the region where the probabilities are
substantially nonzero, the derivative and the integral of the Airy
function are of the same order: $\int_{\varkappa}^{\infty}
\operatorname{Ai}(\xi) \, d\xi \sim \operatorname{Ai}'(\varkappa)$ if
$ 0 < \varkappa \lesssim 1$. Second,
\begin{eqnarray}
\min \left\{ \frac{2 \eta^{2/3}}{ (1-\eta)^{2/3} } + \frac{ (1-
\eta)^{4/3} }{ \eta^{1/3}} \right\} \approx 2, \\
\min \left\{ \frac{1}{ \eta^{1/3} (1-\eta)^{1/3}} - 2\eta^{2/3}
(1-\eta)^{2/3} \right\} \approx 0.8,
\end{eqnarray}
if $\eta \in (0,1)$. Hence, as soon as $\chi \gg 1$, the terms in Eqs.~\eqref{w_basic} and
\eqref{tildew_basic} that are proportional to the integrals of the
Airy function can be neglected. Furthermore, $\varkappa$ is much
smaller than unity on a wide interval $1/\chi' \ll \eta \leq 1$, and
$\tilde \varkappa \ll 1$ if $1/\chi' \ll \eta$ and $1-\eta \gg 1/\chi'$.  Then in
Eqs.~\eqref{w_basic} and \eqref{tildew_basic} we can set $\varkappa =
\tilde \varkappa = 0$.
However, this lead to appearing of the singularities in the expressions for $w$ and
$\tilde w$, namely $w \propto 1/\eta^{1/3} $ at $\eta \rightarrow 0$,
$\tilde w \propto 1/\eta^{1/3}$ if $\eta \rightarrow 0$ and $ \tilde w
\propto 1/(1-\eta)^{1/3}$ if $\eta \rightarrow 1$.
Nevertheless, these singularities are integrable and give contributions to
the total probability rates that are about ${\chi'}^{2/3} \gg 1$ times
smaller than the total probability rates.

Finally, for $\chi' \gg 1$ and $|\psi| \ll 1$ we have:
\begin{eqnarray}
w( {\mathbf p}' \rightarrow {\mathbf p} ) = -\operatorname{Ai}'(0)
\frac{ \alpha a_0^{1/3} }{ \varepsilon_l^{1/3} } \frac{ \psi^{2/3} }{
{p'}^{4/3}} \frac{ 1+ \eta^2}{ \eta^{1/3} (1- \eta)^{2/3} },
\label{w_interim} \\ \tilde w({ \mathbf p}' \rightarrow {\mathbf p} )
= -\operatorname{Ai}'(0) \frac{ \alpha a_0^{1/3} }{ \hbar^{1/3} }
\frac{ \psi^{2/3}}{ {p'}^{4/3} } \frac{ \eta^2 + (1-\eta)^2}{
\eta^{1/3} (1-\eta)^{1/3}}, \label{tildew_interim} \\ W(p',\psi) = v
\frac{\alpha a_0^{1/3} }{ \varepsilon_l^{1/3} } \frac{ \psi^{2/3}}{
{p'}^{1/3} }, \label{W_interim}
\end{eqnarray}
and the same expression for $\tilde W$ except substitution $\tilde v$
for $v$, where
\begin{eqnarray}
v = -\operatorname{Ai}'(0) \int_0^1 \frac{1+ \eta^2}{ \eta^{1/3} ( 1- \eta
)^{2/3} } \, d\eta \approx 1.46, \\
\tilde v = -\operatorname{Ai}'(0) \int_0^1 \frac{ \eta^2 +
(1-\eta)^2}{ \eta^{1/3} (1-\eta)^{1/3} } \, d\eta \approx 0.38.
\end{eqnarray}
Furthermore, Eq.~\eqref{gph_basic} 
can be used in the same form and
Eq.~\eqref{ge_basic} can be simplified as follows:
\begin{multline}
\frac{ \partial g_{e,p} }{ \partial t} = -\frac{ \partial g_{e,p} }{
\partial \psi } \pm \frac{\psi }{ p} \frac{ \partial g_{e,p}}{
\partial \psi } \mp p \frac{\partial}{ \partial p} \frac{ g_{e,p}}{ p}
+ \\ \int_p^{\infty} g_{\gamma}(p', \psi) \tilde w (p'
\rightarrow p, \psi) \, dp' + \\ \int_p^{\infty}
g_{e,p}(p',\psi) w(p' \rightarrow p,\psi) \, dp' - Wg_{e,p}.
\label{ge_interim}
\end{multline}
Similar equations can be written in the region $\chi \gg 1$, $|\psi-\pi| \ll 1$.

The substitution:
\begin{eqnarray}
t = \frac{\varepsilon_l^{1/4}}{ \alpha^{3/4} a_0^{1/4} } \bar t,
\label{bar_t} \\
\psi = \frac{\varepsilon_l^{1/4}}{ \alpha^{3/4} a_0^{1/4} } \bar \psi,
\label{bar_psi} \\
p = \frac{\varepsilon_l^{1/4}}{ \alpha^{3/4} a_0^{1/4} } \bar p.
\label{bar_p}
\end{eqnarray}
in Eqs.~\eqref{gph_basic}, \eqref{w_interim}-\eqref{ge_interim} yields
the following parameter-free equations:
\begin{multline}
\frac{ \partial g_{e,p} }{ \partial \bar t} = -\frac{ \partial g_{e,p} }{
\partial \bar \psi } \pm \frac{\bar \psi }{ \bar p} \frac{ \partial g_{e,p}}{
\partial \bar \psi } \mp \bar p \frac{\partial}{ \partial \bar p} \frac{ g_{e,p}}{ \bar p}
+ \\ \int_{\bar p}^{\infty} g_{\gamma}(\bar p', \bar \psi) \bar
w_{\gamma} (\bar p' \rightarrow \bar p, \bar \psi) \, d\bar p' + \\ \int_{\bar p}^{\infty}
 g_{e,p}(\bar p',\bar \psi) \bar w(\bar p' \rightarrow \bar p,\bar \psi) \, d\bar p' - \bar
Wg_{e,p}, \label{ge_interim_bar}
\end{multline}
\begin{multline}
\frac{\partial g_{\gamma} }{ \partial \bar t} = - \frac{ \partial
g_{\gamma}}{ \partial \bar \psi } - \bar W_{\gamma}
g_{\gamma} + \\ \int_{\bar p}^{\infty} \left[ g_e(\bar p',\bar \psi)+ g_p(\bar p',\bar \psi)
\right] \bar w(\bar p' \rightarrow \bar p'-\bar p, \bar \psi) \, d\bar p',
\label{gph_interim_bar}
\end{multline}
where
\begin{eqnarray}
\bar w( { \bar {\mathbf p}}' \rightarrow {\bar {\mathbf p}} ) =
-\operatorname{Ai}'(0)  \frac{ { \bar \psi}^{2/3} }{ 
{{\bar p}{'}}^{4/3}} \frac{ 1+ \eta^2}{ \eta^{1/3} (1- \eta)^{2/3} },
\label{w_final} \\
\bar w_{\gamma} ({ \bar {\mathbf p}}' \rightarrow {\bar {\mathbf p}} ) =
-\operatorname{Ai}'(0)  \frac{
{\bar \psi}^{2/3}}{ {{\bar p}{'}}^{4/3} } \frac{ \eta^2 + (1-\eta)^2}{
\eta^{1/3} (1-\eta)^{1/3}}, \label{tildew_final} \\
\bar W(\bar p,\bar \psi) = v \frac{
{\bar \psi}^{2/3}}{ {\bar p}^{1/3} }, \label{W_final} \\
\bar W_{\gamma}(\bar p,\bar \psi) = \tilde v \frac{
{\bar \psi}^{2/3}}{ {\bar p}^{1/3} }, \label{W_gamma_final}
\end{eqnarray}
The obtained equations can be used for further analysis of the particle
distribution during the cascade development. However, these equations do
not completely determine the cascade dynamics, because they describe only
the particles with high $\chi$. For example, the particles with $\chi \sim
1$ can also breed and give some contribution to the growth rate.

Let us assume that the growth rate is determined only by the particles with
$\chi \gg 1$, $|\psi| \ll 1$ and $|\psi - \pi | \ll 1$. In this case
the dependence of  $\Gamma$ on the electric field
strength and the wavelength can be determined from Eq.~\eqref{bar_t} as
follows:
\begin{equation}
\frac{ \Gamma(\mu_1, \lambda_{l,1} )}{ \Gamma( \mu_2,\lambda_{ l,2} )}
= \frac{ \mu_1^{1/4} \lambda_{l, 1}^{1/2}}{ \mu_2^{1/4}
\lambda_{l,2}^{1/2} },  \label{Gamma_1/4}
\end{equation}
where $\mu = E_0/\alpha E_{cr}$. The dependence Eq.~\eqref{Gamma_1/4}
and the scaling Eqs.~\eqref{bar_t}-\eqref{bar_p} was also derived in
Ref.~\cite{FedotovArxiv} via simple estimations and dimensional
analysis of the kinetic equations.
However, the numerical simulations show that in a wide range of the parameters
Eq.~\eqref{Gamma_1/4} is not valid (see Sec.~VII and right plot in
Fig.~7 in Ref.~\cite{FedotovArxiv}) and can be used only
for estimations with accuracy about 50\%. Thus, the growth rate
$\Gamma$ is not completely determined by
Eqs.~\eqref{ge_interim_bar}-\eqref{W_gamma_final} that describe only
particles with $\chi \gg 1$. However, the growth rate contributes to
these equations through time derivatives.
Hence, for further analysis of
Eqs.~\eqref{ge_interim_bar}-\eqref{gph_interim_bar} we assume that
$\Gamma$ is known (for example, from the numerical simulations).

\begin{figure}
\includegraphics[]{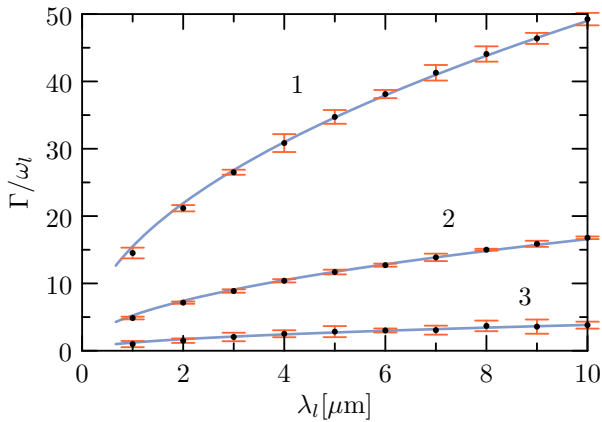}
\caption{The dependence of the growth rate on $\lambda_l$: the numerical
data (dots) and the fits $\Gamma/\omega_l = A \sqrt{\lambda_l}$ for
the data (lines). The amplitude $A$ is chosen such that the RMS deviation
from the numerical data is minimal. The error bars show the dispersion of
the numerical data for 7 runs of the simulation. Line 1
corresponds to the data obtained for $I=10^{29}\text{ W}/\text{cm}^2$, line 2
corresponds to $I=10^{27}\text{ W}/\text{cm}^2$ and line 3
corresponds to $I=10^{25}\text{ W}/\text{cm}^2$.}
\label{fig_gamma-lambda}
\end{figure}

\begin{figure}
\includegraphics[]{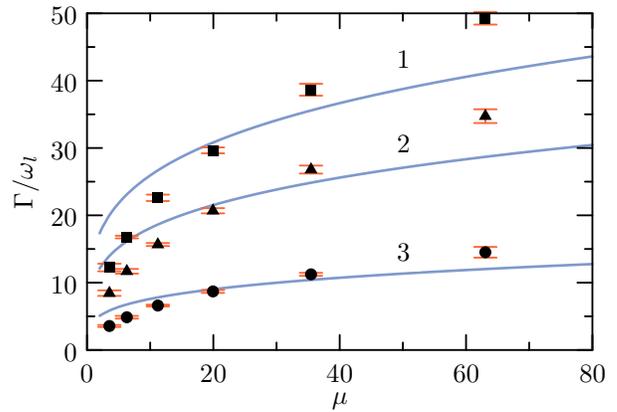}
\caption{The dependence of the growth rate on $\mu = E_0/\alpha
E_{cr}$. The numerical results are presented by the solid black dots. The error bars show the
dispersion of the numerical data for 7 runs of the simulation. The solid lines represent fits
$\Gamma = A \mu^{1/4}$. The amplitude $A$ is chosen such that the RMS
deviation from the numerical data is minimal. Line 1 corresponds to
numerical data for $\lambda_l=10 \text{
}\mu\text{m}$ (squares), line 2 corresponds to $\lambda_l=5 \text{ }\mu\text{m}$ (triangles)
and line 3 corresponds to $\lambda_l=1 \text{ }\mu\text{m}$ (circles).}
\label{fig_gamma-mu}
\end{figure}

\section{Analytical solutions and energy spectra}

Here we again restrict ourself to the limit  $E_0 \gg \alpha E_{cr}$,
and consider the "tails" of the distribution functions that correspond
to $\chi \gg 1$ and, additionally, to $p \gg \langle p \rangle$
in order to estimate the terms in Eqs.~\eqref{ge_interim_bar}, \eqref{gph_interim_bar} and obtain
approximate solutions for the stationary particle distribution functions as well as
for the energy spectra of the electrons, the positrons and the photons. Here $\langle
p \rangle $ is the average particle momentum and the stationary
distribution means that the distribution function shape is approximately 
conserved during the evolution, but the number of particles grows
exponentially:
\begin{eqnarray}
g_{e,p,\gamma} \propto \exp( \Gamma t), \\
\frac{\partial g_{e,p,\gamma}}{ \partial t} = \Gamma g_{e,p,\gamma}.
\end{eqnarray}
The numerical simulations shows that the distribution functions approach
such dependence in time interval about some $1/\Gamma$ after the cascade starts.

The main source of the electrons with  $|\psi| \ll 1$, $p \gg \langle p \rangle
$ is their transfer from the region $p \lesssim \langle p
\rangle$ due to the acceleration. So, electrons are concentrated along
the separatrix $\psi \approx p/2$ owing to the phase flow. We will see
that the $\psi$-width of the electron distribution function at $p \gg
\langle p \rangle $ is much smaller than the angular width of the
photon distribution function and much smaller than $p$. Because of that we
derive the electron energy spectrum and do not derive the angular shape of
the electron distribution function.

Furthermore, we will show that
particle spectra
\begin{equation}
h_{e,p,\gamma}(p) = \int_0^{2 \pi} g_{e,p,\gamma}(p,\psi) \, d\psi
\end{equation}
decrease exponentially with the increase of $p$ due to transfer of
the electrons with high momenta to the region of lower momenta because of 
frequent emission of the photons.  Besides, the characteristic scale
of the decrease is less or about the average momentum $\langle p \rangle$.
Hence, the electrons that acquire momentum $p$ due to the photon
emission mostly have momentum $p'$ before the emission in the
following narrow interval: $p<p' \lesssim p + \langle p \rangle$. The electrons that initially
have momentum $p$ are distributed on wide interval
$p' \in (0,p)$ with the average momentum
about $p$ after the photon emission, because the
characteristic $p'$-scale of the function $w( {\mathbf p} \rightarrow
{\mathbf p}' )$ is the same order as $p$.  The last two terms in
Eq.~\eqref{ge_interim_bar} describe the photon emission. The last but one term in
Eq.~\eqref{ge_interim_bar} describes the inflow of the electrons from
the region of high momenta $p'>p$, and the last term describes the
outflow of the electrons to the region of momenta less than $p$. Summarizing,
we obtain that the last but one term is about $p/\langle p \rangle \gg
1$ times smaller than the last term and can be described in the
framework of the perturbation theory. It can be shown by similar
reasoning that
the term with $g_{\gamma}$ in Eq.~\eqref{ge_interim_bar} can be
neglected.

Integrating Eq.~\eqref{ge_interim_bar} for the electrons over the
interval $\bar \psi \in (\bar p/2 - \Delta \bar \psi, \bar p/2 + \Delta \bar \psi)$, where
$\Delta \bar \psi \ll \bar p/2$ and much higher than the $\bar \psi$-width of
the electron distribution function, and neglecting $g_{\gamma}$, we
obtain:
\begin{multline}
\frac{ \partial h_e }{ \partial \bar p} = - \bar \Gamma h_e - \frac{v {\bar
p}^{1/3} }{ 2^{2/3} } h_e \\- 2^{1/3} \operatorname{Ai}'(0) {\bar p}^{1/3}
\int_0^{\infty} \frac{ h_e( \bar p[1+\xi] ) }{ \xi^{2/3} } \, d\xi,
\label{dhedp}
\end{multline}
where we simplify the formula for $\bar w(p' \rightarrow p, \psi)$ taking into account that $\psi \approx p'/2$ and
$p'-p \sim \langle p \rangle \ll p$. As it was already mentioned, the
characteristic scale of the integrand in the last term of Eq.~\eqref{dhedp} is $\xi \sim \langle p \rangle /p \ll 1$
and this term can be treated as perturbation. In the 
zeroth order of the perturbation theory we have:
\begin{equation}
h_e(\bar p) = h_0 \exp \left[ -\bar \Gamma \bar p - \frac{3v {\bar p}^{4/3} }{
2^{8/3} } \right], \label{he0}
\end{equation}
where $h_0$ is a constant. Then, using the following approximation
derived by the expansion of the terms in the exponent in Tailor series
\begin{equation}
h_e(\bar p [1+ \xi] ) \approx h_e( \bar p) \exp \left[ -\left( \bar \Gamma
\bar p + \frac{ v {\bar p}^{4/3}}{ 2^{2/3} } \right) \xi \right].
\end{equation}
we obtain:
\begin{equation}
\frac{1}{h_e} \frac{ \partial h_e }{ \partial \bar p} = -\bar \Gamma -
\frac{v {\bar p}^{1/3}}{ 2^{2/3} } - \frac{2^{1/3} \operatorname{Ai}'(0)
b_{-2/3} }{ \left[ \bar \Gamma + v {\bar p}^{1/3}/2^{2/3}
\right]^{1/3} }, \label{dhedp1}
\end{equation}
where
\begin{equation}
b_i = \int_0^{\infty} \xi^i \exp(-\xi) \, d\xi,
\end{equation}
and $b_{-2/3} \approx 2.68$. Finally, from Eq.~\eqref{dhedp1} we have:
\begin{multline}
h_e( \bar p) = h_0 \exp\left\{ -\bar \Gamma \bar p - \frac{ 3v {\bar p}^{4/3}}{
2^{8/3} } - \frac{ 9 \operatorname{Ai}'(0) b_{-2/3} {\bar \Gamma}^{8/3} }{ 5
\times 2^{2/3} v^3 } \times \right. \\ \left. \left[ \left( {\mathfrak p}^{1/3} +1 \right)^{2/3}
\left( 5 {\mathfrak p}^{2/3} - 6 {\mathfrak p}^{1/3} + 9 \right) - 9
\right] \right\}, \label{he_final}
\end{multline}
where $\mathfrak p = v^3 \bar p /4 {\bar \Gamma}^3$.

The positrons are decelerated in the region $|\psi | \ll 1$, so, the
only source of the photons in the region $|\psi | \ll 1$, $p \gg
\langle p \rangle $ are the electrons, hence, we can neglect $g_p$ and
assume $\bar p' - \bar p \ll \bar p'$ in Eq.~\eqref{gph_interim_bar}, that yield:
\begin{equation}
\frac{ \partial g_{\gamma} }{ \partial \bar \psi} = -\bar \Gamma g_{\gamma}
- \frac{\tilde v {\bar \psi}^{2/3} }{ {\bar p}^{1/3}} g_{\gamma} - \frac{ 2^{1/3}
  \operatorname{Ai}'(0) h_e(2 \bar \psi ) }{ {\bar p}^{1/3} \left( 2 \bar \psi -
  \bar p \right)^{1/3} }. \label{gph_interim_bar_2}
\end{equation}
In order to obtain this equation we also use the electron distribution
function in the following form:
\begin{equation}
g_e(\bar p', \bar \psi) = h_e(\bar p')
\delta( \bar \psi - \bar p'/2),
\end{equation}
where $\delta(x)$ is the Dirac delta function.
Obviously, the photons, emitted by the electrons in the region $|\psi |
\ll 1$, are situated in the $p-\psi$ space  above the line $\psi =
p/2$. As far as we assume that the photons decay before they pass the
distance about $\pi$ along $\psi$-direction, we can set
$g_{\gamma}(\bar p, \bar p/2) = 0$. Additionally, making of use  the following
approximations in the solution of Eq.~\eqref{gph_interim_bar_2}
\begin{eqnarray}
h_e( 2 \bar \psi) \approx h_e(\bar p) \exp \left[ - \left( \bar \Gamma +
\frac{v {\bar p}^{1/3}}{ 2^{2/3}} \right) ( 2\bar \psi - \bar p)
\right], \label{he2psi} \\
{\bar \psi}^{5/3} - \left( \frac{\bar p}{2} \right)^{5/3} \approx \frac
{5 {\bar p}^{2/3} }{ 3 \times 2^{2/3} } \left( \bar \psi - \frac{ \bar
p}{2} \right), \label{psi53}
\end{eqnarray}
we obtain for $\bar p/2 < \bar \psi \ll 1$:
\begin{multline}
g_{\gamma}(\bar p, \bar \psi) = -\frac{ \operatorname{Ai}'(0) h_e(\bar
p) }{ {\bar p}^{1/3} \left[ \bar \Gamma + \left( 2^{1/3} v - \tilde
v/2^{2/3} \right) {\bar p}^{1/3} \right]^{2/3}} \\ \times \int_0^{\varphi} \frac{
\exp(-\xi) }{ \xi^{1/3}} \, d\xi \\ \times
\exp \left[ -\bar \Gamma \left(
\bar \psi - \frac{ \bar p}{ 2} \right) - \frac{ 3 \tilde v}{ 5 {\bar
p}^{1/3}} \left( {\bar \psi}^{5/3} - \frac{ {\bar p}^{5/3}}{ 2^{5/3} }
\right) \right], \label{gph_final}
\end{multline}
where
\begin{equation}
\varphi = \left[ \bar \Gamma + \left( 2^{1/3} v - \tilde v/2^{2/3} \right)
{\bar p}^{1/3} \right] \left( \psi - \bar p/2 \right).
\end{equation}

\begin{figure}
\includegraphics[]{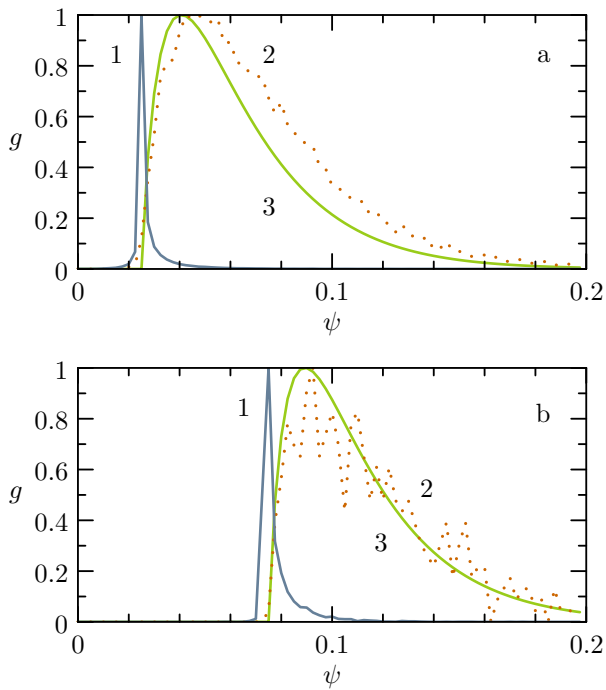}
\caption{The electron (line 1) and the photon (line 2) distribution
functions obtained by the numerical simulation of the
electromagnetic cascade developing in the rotating electromagnetic
field with $a_0 = 8 \times 10^5$ and $\lambda_l = 0.8 \text{
}\mu\text{m}$ for $\gamma$: $\gamma/a_0 = 0.05$ (a)
and $\gamma/a_0 = 0.15$ (b). Lines 3 correspond to the analytical
expression for the photon distribution function Eq.~\eqref{gph_final}. All
plotted functions are normalized to their maximal values.}
\label{fig_article-g-psi}
\end{figure}

\begin{figure}
\includegraphics[]{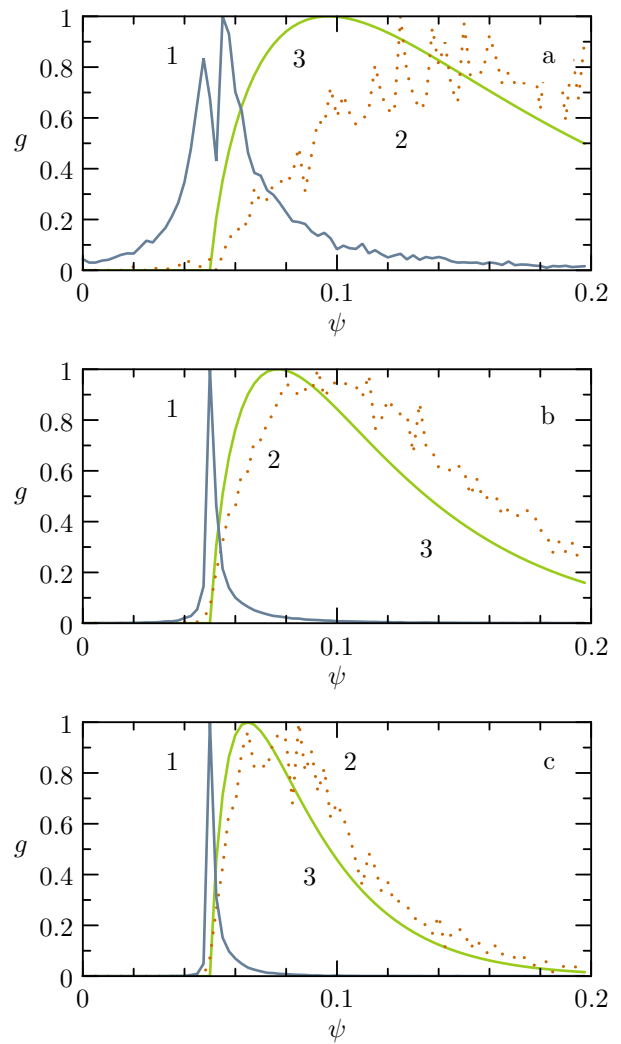}
\caption{The electron (line 1) and the photon (line 2) distribution
functions at $\gamma/a_0 = 0.1$ obtained by the numerical simulations of the
electromagnetic cascade developing in the rotating electromagnetic
field with $\lambda_l = 0.8 \text{ }\mu\text{m}$ and (a) $a_0 = 3.2 \times 10^4$, (b) $a_0 = 1.6 \times 10^5$,
(c) $a_0 = 8 \times 10^5$. Lines 3 correspond  to the analytical
expression for the photon distribution function Eq.~\eqref{gph_final}. All
plotted functions are normalized to their maximal values.}
\label{fig_article-g-psi-a0}
\end{figure}

\begin{figure}
\includegraphics[]{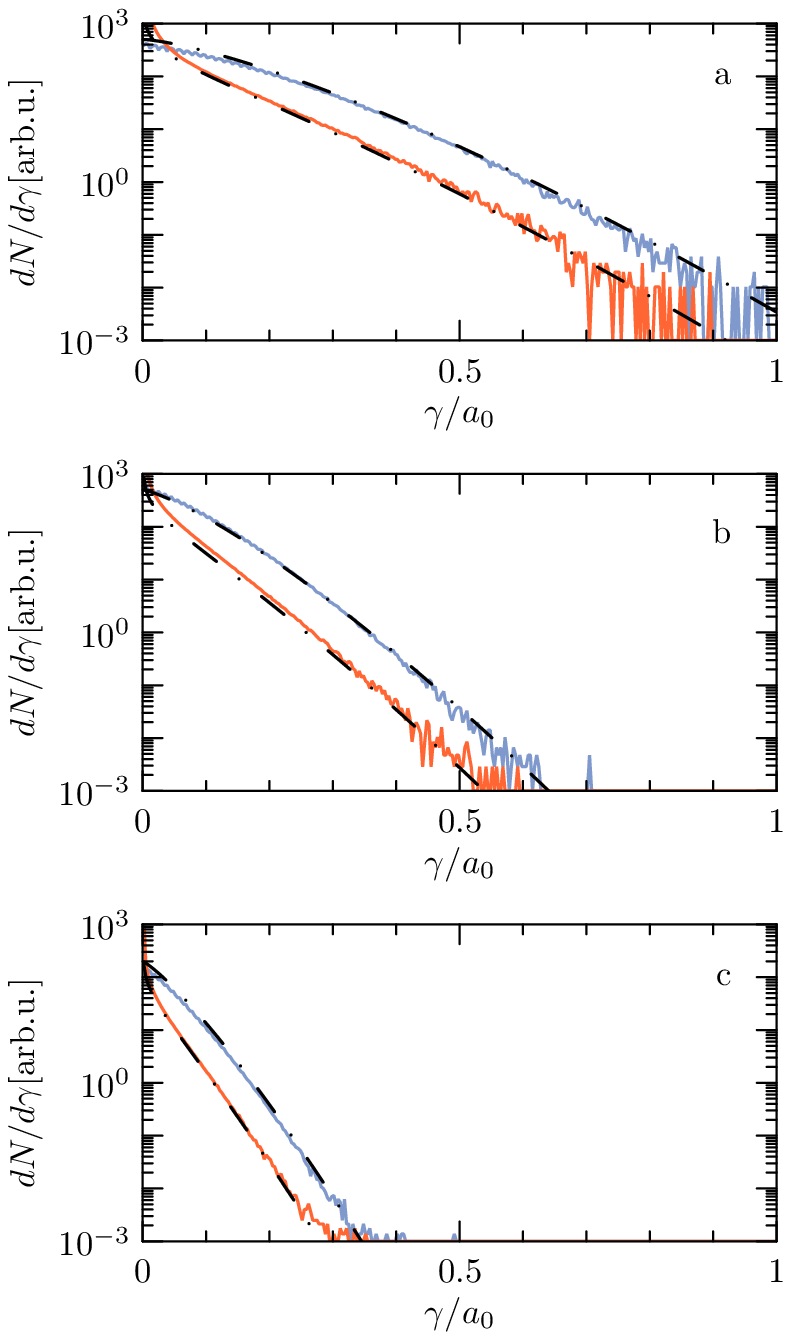}
\caption{The electron (upper lines) and the photon (lower lines) energy
spectra for the electromagnetic cascade developing in the rotating
electric field with $\lambda_l=0.8 \text{ }\mu\text{m}$ and (a) 
$a_0=3.2\times 10^4$, (b) $a_0 = 1.6 \times 10^5$, (c) $a_0 = 8 \times
10^5$. Solid lines correspond to the numerical results and the dash-dotted lines
correspond to the analytical expressions~\eqref{he_final} and
\eqref{hph_final}.}
\label{fig_article-h}
\end{figure}

Further simplifications can be made if we assume 
 that $2^{1/3} v {\bar p}^{1/3} \gg \bar \Gamma$ and take into account that
 $\tilde v /v \approx 0.26 \ll 1$. In this
case the characteristic scale of  the integral in
Eq.~\eqref{gph_final} is much smaller than the scale of the exponent
in Eq.~\eqref{gph_final}. Thus, integrating Eq.~\eqref{gph_final} over
$\psi$, we assume that the integral in this equation is equal to
its maximal value $b_{-1/3}$ if $\varphi > 0$. We also suppose that
the positron distribution function concentrated generally near $\psi =
\pi$ gives the same photon distribution as Eq.~\eqref{gph_final}
except the substitution $\psi-\pi$ for $\psi$. Furthermore, in order
to advance simplification  we can use Taylor's
expansion in the exponent of Eq.~\eqref{gph_final} similarly to
Eqs.~\eqref{he2psi} and \eqref{psi53}. Finally we obtain the following
expression for the photon energy spectrum:
\begin{multline}
h_{\gamma}(\bar p) = -\frac{ 2 \operatorname{Ai}'(0) b_{-1/3} h_e(\bar
p) }{ {\bar p}^{1/3} \left[ \bar \Gamma + \left( 2^{1/3} v - \tilde
v/2^{2/3} \right) {\bar p}^{1/3} \right]^{2/3} } \\ \times \frac{1}{ \left[ \bar \Gamma + \tilde v
{\bar p}^{1/3}/2^{2/3} \right]}. \label{hph_final}
\end{multline}

\section{Numerical simulations}

In this section we compare the analytical expressions for the electron
energy spectrum~\eqref{he_final}, the photon energy
spectrum~\eqref{hph_final} and for the photon distribution
function~\eqref{gph_final} with the results obtained using
the numerical model described in Ref.~\cite{PRL2011}.
Furthermore, using the numerical values of $\bar \Gamma$, we compare
various terms in these expressions and simplify them for a wide
range of parameters.

The electron and the photon distribution functions in electromagnetic
cascade are shown in Fig.~\ref{fig_spectrum-g}. These functions are
obtained in the numerical simulation performed for the field parameters
$a_0 = 1.6 \times 10^5$, $\lambda_l = 0.8 \text{ }\mu\text{m}$.
The numerical simulation gives $\Gamma/\omega_l \approx 12.1$ for
these parameters. The shape of the distribution functions is typical for
the cascades with $\Gamma \gg 1$. The electron distribution function
at $p \gg \langle p \rangle$ is concentrated near the line $\psi =
p/2$, and the photon distribution function is nonzero upper this line,
that agrees well with the qualitative description given in Sec.~VI.
Furthermore, it follows from Fig.~\ref{fig_spectrum-g} and the results of
the numerical simulation that almost all particles are located
in the regions $p \ll 1$, $|\psi| \ll 1$ and $p \ll 1$, $|\psi - \pi|
\ll 1$.

The dependence of the growth rate $\Gamma$ on the electric field
parameters is shown on Fig.~\ref{fig_gamma-surface-plot}. In every
node of the surface $\Gamma$ is calculated $7$ times and the result
is averaged out. It is seen that $\Gamma$ grows with the increase of
the intensity and the wavelength, and the growth of
$\Gamma$ due to the wavelength increase  is much steeper than the
growth due to the intensity increase. Therefore, it is preferable
to use longer wavelengths in the experiments devoting to the cascades
observation. Besides, the limitation of the attainable intensity
due to prolific pair production is less strict for shorter
wavelengths.

In Fig.~\ref{fig_gamma-lambda} the dependence of the growth rate
on the wavelength for three different values of the intensity is shown. The
rows of dots show the results of the numerical simulations. The error bars show
the dispersion of $\Gamma$ obtained for $7$ simulations with the same
initial parameters.  The cascades in the simulations are initiated by
a single immobile $e^+e^-$ pair. Thus, at initial stage of the cascade
development when the number of cascade particles is small the shape
of the distribution functions is being modified. Because of this the
growth rate fluctuates significantly at the initial stage. We
measure the growth rate as $\Gamma = d( \ln N_e )/dt$ at the later stage
of the cascade development when the number of particles is high and
the growth rate becomes constant. The dispersion of $\Gamma$,
obtained in the simulations, is explained by the finiteness of the particle number
used in the simulations. The solid lines in
Fig.~\ref{fig_gamma-lambda} are the fits described by the equation
$\Gamma = A \sqrt{\lambda_l}$, where $A$ is chosen for every row of
the numerical data such that the RMS difference between the numerical data and
the fit is minimal. The line 1 corresponds to the numerical data obtained for
$I = 10^{29} \text{ W}/\text{cm}^2$, the line 2 corresponds to the
numerical data obtained for $I = 10^{27} \text{ W}/\text{cm}^2$ and
the line 3 corresponds to the numerical data obtained for $I = 10^{25}
\text{ W}/\text{cm}^2$. It is seen from Fig.~\ref{fig_gamma-lambda}
that the theoretically obtained dependence of the growth rate on the
wavelength Eq.~\eqref{Gamma_lambda} is in good agreement with the results of
the numerical simulations.

The dependence of the growth rate on the electric field strength is
shown in Fig.~\ref{fig_gamma-mu}. Black points present the results of
the numerical simulations, and the solid lines are the RMS fits $\Gamma = A
\mu^{1/4}$ that correspond to Eq.~\eqref{Gamma_1/4}. The line 1 corresponds to
the squares that illustrate the numerical results obtained for $\lambda_l = 10
\text{ }\mu\text{m}$, the line 2 belongs to the triangles that illustrate
the numerical data for $\lambda_l = 5 \text{ }\mu\text{m}$ and the line 3
corresponds to the numerical data obtained for $\lambda_l = 1 \text{ }\mu
\text{m}$ (circles). It follows from Fig.~\ref{fig_gamma-mu} that
Eq.~\eqref{Gamma_1/4}  can be
used only for estimations of $\Gamma$ with accuracy about $50\text{\%}$.
However, Eq.~\eqref{Gamma_1/4} is derived under the assumption that
$\chi \gg 1$ for the most of cascade particles, but actually many cascade
particles can have $\chi \sim 1$. These particles can give substantial
contribution to the cascade dynamics and, hence, to the cascade growth
rate. Indeed, numerical simulations show that
for $I= 10^{27} \text{ W}/\text{cm}^2$, $\lambda_l =
1\text{ } \mu \text{m}$ the portion of the electrons with $\chi < 5$ is
about $30\text{\%}$ and this portion for $I= 10^{25} \text{ W}/\text{cm}^2$, $\lambda_l =
1\text{ } \mu \text{m}$ is about $90 \text{\%}$.

The comparison between Eq.~\eqref{gph_final} and  $g_{ph}(\psi)$ at certain value of $p$ obtained by
the numerical simulations is given in Fig.~\ref{fig_article-g-psi}. The line 1
and the line 2 correspond to the electron and the photon distribution functions,
respectively. The line 3 corresponds to the solution~\eqref{gph_final}. The
 distribution functions are normalized to their maximal values.
The parameters of the electric field are: $a_0 = 8\times
10^5$, $\lambda_l = 0.8 \text{ }\mu\text{m}$. The distribution
functions $g_{e,\gamma}(\psi,p)$ are shown at $p= 0.05$ (plate a) and
at $p=0.15$ (plate b). The growth rate used in Eq.~\eqref{gph_final}
is calculated numerically: $\Gamma/\omega_l = 24.1$. It is seen that the electron distribution
function is much narrower than that of the photons.
Furthermore, the numerical results are in good agreement with
Eq.~\eqref{gph_final}. Similar cuts of the distribution functions are
shown in Fig.~\ref{fig_article-g-psi-a0}. The panel (a) corresponds to $a_0 = 3.2
\times 10^4$ and $\Gamma = 6.29$ obtained in the simulation, the panel (b)
corresponds to $a_0 = 1.6 \times 10^5$ and $\Gamma = 12.1$, the panel (c)
corresponds to $a_0 = 8 \times 10^5$ and $\Gamma = 24.1$. It is seen
from the plates (b) and (c) that Eq.~\eqref{gph_final} is in good
agreement with the results of the numerical simulation. However, the
photon distribution function obtained theoretically reaches its
maximum earlier than the photon distribution function obtained in
numerical simulations. Possible explanation of this small discrepancy
is the following. In order to obtain Eq.~\eqref{gph_final} we assume
that the electron distribution function depends on the angle $\psi$ as
follows: $g_e(p,\psi) \propto \delta( \psi - p/2 )$, however,
the numerical simulations show that $\psi$-width of the electron
distribution function is finite and big amount of electrons have
$\psi$ slightly higher than $p/2$, hence, the maximum of the photon distribution
function is also slightly shifted to the region of higher angles.
Besides, for the plate (a) the condition $p \gg \langle p \rangle$ is
not fulfilled, the shape of electron distribution function in this
case becomes more complex than in the case $p \gg \langle p \rangle$
and Eq.~\eqref{gph_final} gives rather qualitative approximation for
the photon distribution function.

In Fig.~\ref{fig_article-h} the electron and the photon energy
spectra are shown. The solid lines correspond to the results of
the numerical simulations and the dash-dotted lines correspond to
Eqs.~\eqref{he_final} and \eqref{hph_final}. The top lines correspond to
the electron distribution functions and the bottom lines correspond to
the photon distribution functions. In Eqs.~\eqref{he_final} and
\eqref{hph_final} we use $\Gamma$ from numerical simulations.
$\lambda_l = 0.8 \text{ } \mu \text{m}$, plate (a): $a_0 = 3.2 \times
10^4$, $\Gamma = 6.29$, plate (b): $a_0 = 1.6 \times 10^5$, $\Gamma =
12.1$, plate (c): $a_0 = 8 \times 10^5$, $\Gamma = 24.1$. It is seen
that the obtained analytical results are in good agreement with the results of
the numerical simulations.

\section{Further simplifications}

The obtained analytical expressions for the particle distribution
functions are
quite complex. Besides, these expressions contain the growth rate
$\Gamma$ that is not derived in the framework of the presented theory.
In this section we compare various terms in the obtained
expressions and
derive asymptotic ($p \rightarrow \infty$) expressions for the
particle spectra that are more simple than Eqs.~\eqref{he_final},
\eqref{hph_final}. In order to obtain these asymptotic spectra we
keep in Eqs.~\eqref{he_final} and \eqref{hph_final} only the terms
that describe the slope of the logarithmic spectra and the ratio between the
electron and the photon energy spectra. The slope and the ratio can be probably
measured in the future cascade experiments that allows estimating of
the field magnitude in the cascade region and the growth rate. Furthermore, we
find the fit for $\Gamma$ guided by the results of the
numerical simulations in a wide range of the parameters ($\lambda_l = 1 -
10 \text{ }\mu \text{m}$, $I = 10^{25}-10^{29} \text{
W}/\text{cm}^2$).

\begin{figure}
\includegraphics[]{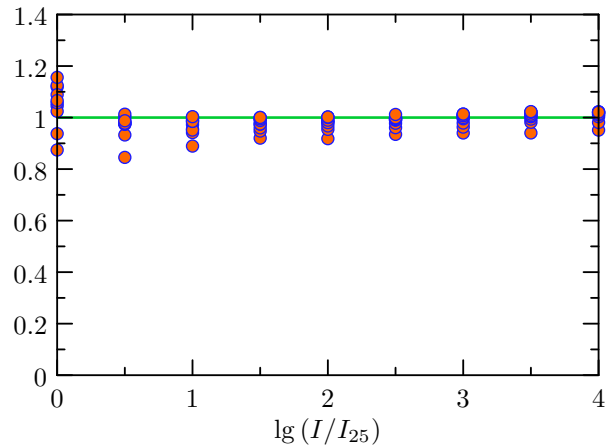}
\caption{The ratio of the growth rate, $\Gamma$, to the fit
Eq.~\eqref{gamma-fit} (dots) and unity level (line). $\Gamma$
is computed for the intensity range $I = 10^{25} - 10^{29} \text{
W}/\text{cm}^2$ and the range of wavelengths $\lambda = 1 - 10 \text{
} \mu\text{m}$ with the step $1 \text{ }\mu\text{m}$.}
\label{fig_gamma-fit}
\end{figure}

\begin{figure}
\includegraphics[]{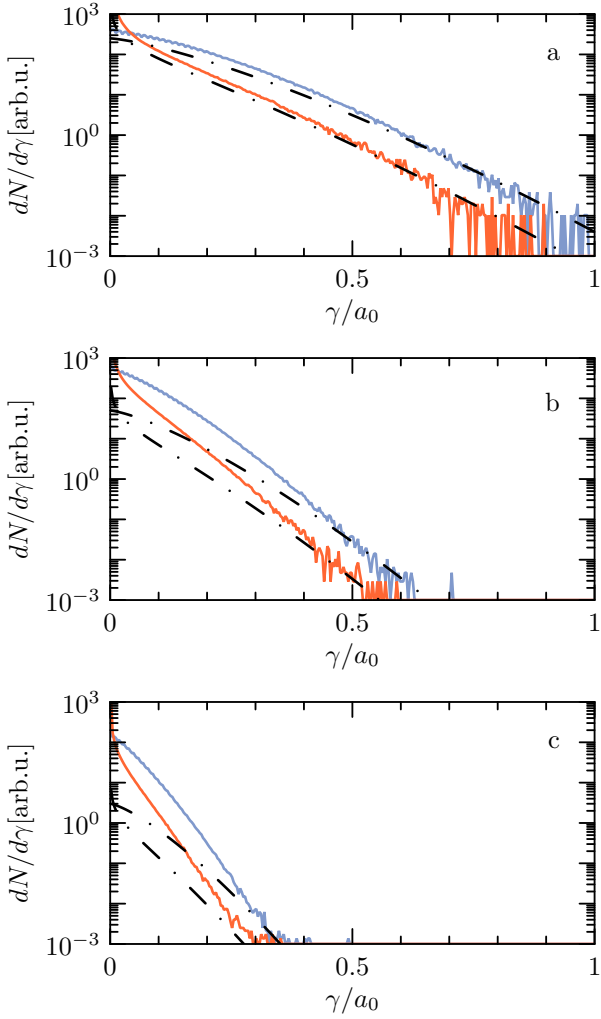}
\caption{The electron (upper solid lines) and the photon (lower solid
lines) energy spectra obtained by the numerical simulations of
the electromagnetic cascade developed in the field characterized by
$\lambda = 0.8 \text{ }\mu \text{m}$ and $a_0 = 3.2 \times 10^4$
(plate (a)), $a_0 = 1.6 \times 10^5$ (plate (b)), $a_0 = 8 \times
10^5$ (plate c). The corresponding asymptotic expressions~\eqref{hes}
and \eqref{hphs} are shown by the dash-dotted lines.}
\label{fig_article-h-simplified}
\end{figure}

We can propose the following fit for the growth rate:
\begin{equation}
\Gamma/\omega_l = 0.65 \times \lambda_l^{1/2}[\mu\text{m} ] \left(
\frac{I}{I_0} \right)^{1/8} \lg \frac{I}{I_0}, \label{gamma-fit}
\end{equation}
where $I_0 = 6\times 10^{23}\text{ W}/ \text{cm}^2$, the magnitude and
$I_0$ was chosen in order to obtain good agreement with $\Gamma$ from
the numerical simulations. The ratio of $\Gamma$ obtained in the numerical
simulations of Fig.~\ref{fig_gamma-surface-plot} to the fit Eq.~\eqref{gamma-fit} (circles) is shown
in Fig.~\ref{fig_gamma-fit}. It is
seen that Eq.~\eqref{gamma-fit} has a satisfactory accuracy.

Making of use Eq.~\eqref{gamma-fit} we can estimate the terms in
Eqs.~\eqref{he_final}, \eqref{gph_final} and \eqref{hph_final}.
First, for $\bar \Gamma$ we obtain the following approximation:
\begin{equation}
\bar \Gamma \approx 0.22 \lg \frac{I}{I_0}.
\end{equation}
Thus, the ratio of the first and the second terms in the exponent in
Eq.~\eqref{he_final} is
\begin{equation}
\frac{3 v {\bar p}^{4/3} }{ 2^{8/3} \bar \Gamma \bar p} \approx \frac{ 3\times{\bar
p}^{1/3} }{ \lg I/I_0 }.
\end{equation}
For the considered region of parameters $\lg I/I_0 \approx 1 - 5$.
Therefore, asymptotically at high values of $\bar p$ this ratio is much
more than unity and we can neglect the first term in the exponent in
Eq.~\eqref{he_final}.
Besides, the last
term in the exponent in Eq.~\eqref{he_final} relates to the second term as
follows,
\begin{multline}
\frac{0.3}{ {\bar \Gamma}^{4/3} \mathfrak p^{4/3} } \left[ \left( \mathfrak
p^{1/3} + 1 \right)^{2/3} \right. \\ \times \left. \left( 5\mathfrak p^{2/3} - 6 \mathfrak
p^{1/3} + 9  \right) - 9 \right],
\end{multline}
where $\mathfrak p$ is introduced right after Eq.~\eqref{he_final}.
This expression decreases with the increase of $\mathfrak p \approx 0.8
\bar p/ {\bar \Gamma}^3$, hence, for sufficiently large values of
$\bar p$ the last term can be neglected for
simplification. Therefore, the electron spectrum
Eq.~\eqref{he_final} can be reduced to the following asymptotic
form:
\begin{multline}
h_{e,a}(\gamma) = h_0 \exp \left[ 
- 0.69 {\bar p}^{4/3} \right] \\ = h_0 \exp \left[ -\gamma^{4/3}/\gamma_*^{4/3}
  \right],
\label{hes}
\end{multline}
where $\bar p = \alpha^{3/4} \varepsilon_l^{-1/4} a_0^{-3/4} \gamma \approx
1.3 \gamma/\gamma_*$ and
\begin{equation}
\gamma_*= 2.4 \times
10^{-7} \lambda_l^{1/2} [ \mu \text{m}] I^{3/8}[\text{W}/\text{cm}^2] .
\end{equation}

In order to simplify Eq.~\eqref{hph_final} we neglect $\tilde v/2
\approx 0.19$ in comparison with $v \approx 1.46$ and we neglect $\bar
\Gamma$ in comparison with
$2^{1/3} \tilde v {\bar p}^{1/3} \approx 1.8 {\bar p}^{1/3}$. However,
we does not neglect $2^{-2/3} \tilde v {\bar p}^{1/3} \approx 0.24
{\bar p}^{1/3}$ in comparison with $\bar \Gamma$. The
resulting asymptotic form of the photon spectrum is
\begin{equation}
h_{\gamma,a}(\gamma) = \frac{ 0.47 h_{e,s}(\gamma) }{ {\bar p}^{5/9}
\left[ \bar \Gamma + 0.24 {\bar p}^{1/3} \right]}, \label{hphs}
\end{equation}
where \begin{multline}
\bar \Gamma = \varepsilon_l^{1/4} \alpha^{-3/4} a_0^{-1/4}
\Gamma/\omega_l \\ \approx \frac{323}{ \lambda_l^{1/2} [\mu \text{m}]
I^{1/8}[\text{W} / \text{cm}^2] } \frac{ \Gamma}{ \omega_l }.
\end{multline}
The comparison of Eqs.~\eqref{hes} and \eqref{hphs}
with the results of the numerical simulations is given in Fig.~\ref{fig_article-h-simplified}.

\section{Summary and discussion}

In this paper we study the development of the electromagnetic cascade
in the rotating electromagnetic field. First we derive the kinetic equations for the electrons, 
the positrons and the photons in the cascade. Second, we present the dimensional analysis 
and scaling for the kinetic cascade equations. In the limit when the growth rate is
quite large $\Gamma \gg 1$ the angle between the momenta of 
the cascade particles and the electric field vector is small. This allows us to
find how the cascade parameters such as the average particle 
momentum and the cascade growth rate depend on the frequency of the
field rotation.
It is shown that lower frequencies is preferable
in the experiments devoting to the cascades
observation. At the same time, the limitation of the attainable intensity
due to prolific pair production is less strict for higher
frequencies.
Then, under the assumption that the most of particles have $\chi \gg
1$ the dependence of the cascade parameters on the electric field
strength is found. Besides, for the
distribution
 function "tails" corresponding to the particles with high values of
 quantum parameter $\chi \gg 1$ and with Lorentz factor $\gamma \gg
 \langle \gamma \rangle$, 
the analytical solution is found. The obtained formulae for the distribution
functions and for the energy spectra agree fairly good with the results of
the numerical simulations.  Furthermore, the fit for the growth rate
based on the results of the numerical simulations in a wide range of
the parameters is proposed.

The electromagnetic cascades can arise in various realistic field
configurations, particularly, in the field of circularly or linearly
polarized standing wave \cite{PRL2011}. In the field of the circularly
polarized standing wave the cascade develops mostly in the $B$-node,
where $B \approx 0$. Near this plane the field configuration is very
close to the homogeneous rotating electric field. Certainly, some
particles can leave the plane $B=0$. However, we assume that this fact
does not affect cascade dynamics much and the proposed theory fits
well for the description of the electromagnetic cascades in the
circularly polarized standing wave. The estimations also demonstrate that the
cascade can arise in the crossed time-dependent electric and magnetic
fields if the magnitude of the electric field is greater than the
magnitude of the magnetic field. In this case particle momenta are also
concentrated along the distinguished direction and some
simplifications of the kinetic equations are also possible. Thus, the
proposed model for the electromagnetic cascades in rotating electric field
can be used for analysis of the cascade dynamics in some other field configurations.

It follows from the numerical simulations that in the considered region of the parameters the big portion of the particles
has $\chi \sim 1$ and can strongly affect the cascade development and the cascade
growth rate. As far as the obtained distribution functions for the particles with
$\chi \gg 1$ depend on the cascade growth rate, the proposed
description of the "tails" of the distribution functions is not
completely
self-consistent. The contribution of the particles with $\chi \sim 1$ 
in $\Gamma $ is especially substantial for the near-threshold cascades. Thus,
the investigation of the distribution function component with $\chi \sim 1$ is
 important for experimental observation electromagnetic cascades in laser field at
the lowest possible intensity level.

However, it is important to note that the predictions of the proposed
theory for the dependence of the cascade parameters on the frequency
of the field rotation and for the distribution functions and the
energy spectra of the high-energy cascade particles are in fairly good
agreement with
the results of the numerical simulations. The proposed model may be
useful for the development of $e^+e^-$ plasma sources and for the
better understanding of some astrophysical processes. Furthermore, the
obtained expressions might be also used to estimate the
cascade parameters in the experiments by measuring the particle energy spectra.

\begin{acknowledgments}
We are grateful to A.~M.~Fedotov and N.~B.~Narozhny for fruitful
discussions.  This work has been supported by federal target program
"The scientific and scientific-pedagogical personnel of innovation in
Russia" and by the Russian Foundation for Basic Research.
\end{acknowledgments}

\end{document}